\documentstyle[aps,epsf]{revtex}
\begin{document}
\draft
\title{Extensional rupture of model non-Newtonian fluid filaments}
\author{Joel Koplik}
\address{Benjamin Levich Institute and Department of Physics,\\
City College of the City University of New York, New York, NY 10031}
\author{Jayanth R. Banavar}
\address{Department of Physics, 
The Pennsylvania State University, University Park, PA 16802}

\date{\today}
\maketitle

\begin{abstract}
We present molecular dynamics computer simulations of filaments of model 
non-Newtonian liquid stretched in a uniaxial deformation to the point of 
breaking.  The liquid consists of Lennard-Jones monomers bound into chains 
of 100 monomers by nonlinear springs, and several different 
constant velocity and constant strain rate deformations are considered.
Generally we observe nonuniform extensions originating in an interplay 
between the stretching forces and elastic and capillary restoring mechanisms, 
leading to highly uneven shapes and alternating stretched and unstretched 
regions of liquid.  Except at the fastest pulling speeds, the filaments
continue to thin indefinitely and break only when depleted of molecules,
rather than common viscoelastic rupture mechanisms.  
\end{abstract}

\section{Introduction}
The extensional dynamics of polymeric, non-Newtonian liquids controls
processes ranging from commonplace activities such as chewing gum to 
large-scale commercial applications in extrusion processing of materials
\cite{engr}.  Understanding the dynamics of the liquid in these situations 
\cite{bird} requires information beyond the usual shear viscosity 
characterization of flowing fluids, and has led to extensive experimental and 
theoretical work on ``liquid bridges'' \cite{ms}.  Here, one pulls on the
ends of a cylinder of the liquid and observes shapes, forces, velocities and
(to some degree) stresses, and attempts to relate the behavior to an
appropriate rheological model.  The ultimate fate of an extending liquid
bridge is a rupture at some point, and one component of the efforts in this
subject has attempted to understand the fracture mechanisms involved.
Aside from practical relevance to materials failure, the thinning dynamics
of a cylindrical neck of liquid has become an important subject in fluid
interfacial dynamics \cite{eggers}.  In almost all cases, these efforts have
concerned macroscopic bodies of liquid, in the sense that the experiments do
not generally have atomic resolution, and the calculations are based on 
continuum descriptions.  The purpose of this paper is to consider the
interfacial rupture of a liquid bridge configuration at an atomic scale.
In addition to the intellectual question of exploring the late stages of 
rupture that are below the level of resolution considered previously, a 
further motivation comes from recent interest in micro- and nano-fluidics 
\cite{nano}, where precisely these length and time scales can be reached in
practice.

The focus of this paper is somewhat outside the issues usually considered in
rheology.  Previous studies of the rupture of liquid filaments \cite{mp}
have concentrated on determining instability criteria based on continuum 
modeling.
The focus has been on constant strain-rate deformations, and on identifying
characteristic regions of the strain rate -- maximum strain phase space,
corresponding to viscous, elastic, rubbery and glassy behavior.  It has proven
useful to distinguish a recoverable component of the strain, representing
the residual deformation of the material when the stretching forces are
released and it is allowed to retract (i.e., omitting viscous flow
deformations).  It is inferred 
that true rupture occurs only when this recoverable strain exceeds 1/2.  
In the systems studied here, the
chain molecules can distort in shape but not divide or combine, but one
possible mechanism for a nonrecoverable strain could involve permanent 
disentanglement of the chains.
A second focus of previous work has been the relation between time to
rupture and applied stress;  in our simulations the ends of the filament are
moved at prescribed velocities, so that applied stress is not
fixed and instead fluctuates during the course of the stretch.  To discuss
other configurations, empirical correlations or the results of non-Newtonian
continuum modeling are used.  The simulations performed here quickly lead to
regions of filament which are only a few molecules in diameter, and not
obviously described in these ways. 

We have previously considered the rupture of fluid interfaces in two
situations, using molecular dynamics (MD) simulations.  In a study of the 
dynamics of a monatomic drop placed in an second 
immiscible fluid undergoing shear \cite{old_rupt}, we found fair
quantitative agreement with experiments and continuum calculations in
general, and in particular that 
the drop ruptures at a sufficiently large capillary number.  The
microscopic mechanism involved is a gradual flow of the drop's molecules 
to one side or another induced by the external flow, coupled to surface 
tension (originating in the molecular interactions chosen to provide 
immiscibility) acting to reduce the area of the interface separating the 
drop and solvent liquids.   In a previous paper on non-Newtonian fluids
\cite{bkb} we used MD simulations to study liquid bridge deformation, 
focusing on the shape evolution and internal velocity and stress fields.
In that paper relatively small samples of several different model molecular
fluids were considered -- a Newtonian liquid based on short chain molecules, 
a solution of a few longer (20-40 monomer) molecules in a short-chain solvent, 
and melts of the longer chains.  Although
the simulations were continued to the point of rupture, these systems were
too small to display any interesting behavior in that regard which could be
quantitatively analyzed.
The Newtonian liquid and melt systems ruptured due to fracture after only 
modest length increases, by less than a factor of two.  The model
solution system attained rather longer lengths, by virtue of the longer
molecules becoming extended and providing a kind of backbone, and ruptured 
when the overall length exceeded the amount of molecular backbone available,
essentially following the monatomic mechanism.
Our goal here is to use a larger system based on a melt of longer molecules,
and study the deformations of highly elongated liquid filaments.

\section{Simulation Procedure}

The simulational method and the molecular system closely follows our earlier
papers on nanodrop dynamics \cite{kpb}, liquid bridge extensional flows
\cite{bkb}, and single chains in a flow \cite{cckb}.  
The MD calculations themselves are of standard form \cite{at,rev}, and 
consider a fluid made of monomers interacting 
with two-body forces of two types, a Lennard-Jones potential to provide an 
impenetrable core and chemical attraction at larger distances, along 
with a finitely-extensible nonlinear elastic force to link the monomers into 
model polymer chains.  The explicit forms of the interaction potentials are
$$
V_{\rm LJ}(r) = 4\epsilon \left[ \left( {r\over\sigma} \right)^{-12} -
\left( {r\over\sigma} \right)^{-6}\ \right]
\qquad 
V_{\rm FENE}(r)={1\over 2} k r_0^2\ \ln\left(1-{r^2\over r_0^2}\right)
\eqno(1)
$$
The Lennard-Jones interaction acts between any two monomers separated by a 
distance less than $r_c=2.5\sigma$, and a linear term is added to $V_{\rm
LJ}$ so that the force vanishes at this cutoff.  The FENE potential 
\cite{bird} acts only between 
adjoining monomers in any one molecule, and has the effect of limiting the 
bond length to $r_0$.  The parameters in the FENE potential
are taken to be $r_0=1.5\sigma$ and $k=30\epsilon/\sigma^2$, following
Kremer and Grest \cite{kg},
and are chosen to yield an entangled melt of non-crossing chains for the 
100-monomer chain length used here.
The characteristic time for monomer dynamics is defined by 
the LJ parameters, $\tau=\sqrt{\epsilon/m\sigma^2}$, where $m$ is the monomer 
mass.  In the remainder of this paper, and in particular in all figures, 
when dimensions are not given we are implicitly using ``MD units'' where 
$\sigma$ is the unit of length, $\tau$ is the unit of time, and $\epsilon$
the unit of energy.  

A molecular liquid based on the FENE and LJ potentials (1)
has been widely studied as an MD a model of polymer melts \cite{binder}.
In most previous work, the 6-12 potential is cut off at a smaller value than
used here, $r_c= 2^{1/6}\sigma$, which has the computational advantage of 
reducing the number of particles which are in interaction range, and speeding
the calculation. The properties of this system are given in detail in
\cite{klh}.  This value of $r_c$ corresponds to a purely repulsive potential
which, while adequate for a confined material, has the disadvantage here that
a fluid will expand to fill space uniformly, with no interface.  Since an
interface is crucial here, attraction is needed and we retain the larger 
value of $r_c$.  However, 
because the interaction has been altered, the liquid studied here need not 
have the same transport coefficients.  In separate simulations we find
a static shear viscosity $\mu(0)\sim 10^3 m/\sigma\tau$, defined as the 
limit of the dynamic shear viscosity $\mu(\dot{\gamma})$,  the
ratio of shear stress to strain rate in Couette flow, when the strain rate 
tends to zero.  Typically polymer melts based on LJ and FENE
potentials show shear thinning -- a nearly constant viscosity up to a
certain transition shear rate $\dot{\gamma}^*$, and a systematic (roughly
power-law) decrease at larger values.  We use the inverse of the transition 
shear rate as a characteristic fluid relaxation time, $\lambda\equiv
1/\dot{\gamma}^*\sim 10^5\tau$, in this case.  This number and the quoted
viscosity above are very crude estimates: at the very low velocities needed 
to reach the plateau for this chain length, the signal-to-noise ratio in a 
stress measurement is poor, and even in very long runs substantial 
fluctuations are found.  The rheology of chain liquids and the definition of
characteristic time in this context is discussed further in \cite{klh,bkb};
the latter paper considers MD simulations of the liquid bridge extensional 
dynamics of model melts (and other fluids) and also shows that they exhibit
strain-hardening in extensional flow, in the sense that an appropriate 
extensional viscosity increases with strain rate.
Lastly, the surface tension is measured to be $\alpha=0.37\pm 0.01$ using an 
independent equilibrium simulation of a slab of liquid with planar interfaces
\cite{widom}.

The MD calculations are performed in parallel on either T-3E, 
Origin-3000 or IBM SP platforms, using a 
spatial domain decomposition algorithm, a layered linked-cell list on each
processor, and MPI routines.  The numerical procedure is to
integrate Newton's equations of motion using the velocity Verlet algorithm
\cite{at} with time step 0.005$\tau$. Occasionally the computation was 
checked by re-running over a certain time interval with a smaller step 
length.  A Nos\'{e}-Hoover thermostat \cite{nh} is used to 
maintain the liquid at a constant temperature $T=1.5\epsilon/k_B$.
The initial configuration, a roughly-circular cylinder of length
$L_0=80.5\sigma$ and radius $R_0=26.3\sigma$, containing 1400
chain molecules 100 monomers in length, is in a disordered liquid state
resulting from a somewhat convoluted preparation procedure.  In our earlier
simulations of length-100 chain molecule melts \cite{kpb}, we began with 
the monomers on regular lattice sites and randomized the configuration 
by cookingit at a high 
temperature and low density in a box with repulsive walls.  This system 
was equilibrated to the point where its time-averaged monomer spatial 
density was uniform, the probability distributions of molecular sizes
(end-to-end length and radius of gyration) reached a steady state, and 
the probability distributions of molecular orientations were rotationally 
invariant.  A spherical drop was formed by cooling the liquid gradually to 
temperature 1.5, while simultaneously applying a weak central force until a
stable sphere formed, and then gradually switching off the force.  Next, half 
of the sphere was placed on a partially-wetting atomistic solid substrate and
re-equilibrated, and finally the sphere was allowed to coalesce with a slightly
randomized copy of itself.  Because both hemispheres were placed on
attracting substrates, the result was a liquid cylinder held between two 
solid surfaces, rather than a single larger sphere.  Removing the solid 
substrate after coalescence gives the initial cylinder considered
here.  One might be concerned that when the cylinder is pulled it would tend
to break at the mid-plane into its two original drops due to incomplete
healing, but in fact we observe rupture at various other locations.   

To stretch the filament, we simply pull on its ends. More precisely, each
monomer in a disk of thickness 5$\sigma$ at each end is attached to a
``tether site'' by a stiff spring, and the tethers are moved outwards in a
prescribed manner by displacing their position by the
appropriate axial amount $\pm v_0 \Delta t\; \hat{z}$ at each time step.  
In simulations of constant-velocity extension, $v_0$ is just the
velocities at each end of the filament, while for constant-strain-rate 
extension where the length of the filament grows as 
$L(t)=L_0e^{\dot{\epsilon}_0 t}$,
the fixed control parameter is the strain rate $\dot{\epsilon}_0$,
and the velocity is $v_0(t)=dL(t)/dt$.   This procedure is similar to a 
common experimental one 
\cite{ms} where the liquid filament rests between solid end plates which are 
moved apart, but has the conveniences that the end plates need not be modeled
and that flow along the end plates does not arise.  The focus of this paper is
the behavior of the liquid in the ``middle''  of the filament, so nothing is
lost in this way.  We also attempted an alternative procedure suggested by
experiments on single DNA molecules \cite{sfb} in which a
constant axial force was applied to each monomer in the end regions, but
the result was that the end regions contracted radially and ruptured occurred
nearby, leaving the bulk of the filament unextended.

The operating conditions of the various simulations performed here are listed 
in Table~I, in terms of the pulling velocity or strain rate. Note that in
either case the initial velocity and strain rate are related by 
$v_0=L_0\dot{\epsilon}_0$, where $L_0$ is the 
filament's initial length.  In terms of the usual dimensionless groups 
characterizing a flow, at the start of a run the Reynolds number 
$Re = \rho v_0 R_0/\mu(0)$ is always much less than unity, while
the Deborah number $De = \lambda\dot{\epsilon}_0$ and the capillary number 
$Ca = \mu(0) v_0/\alpha$ are O($10^2$) or larger.  
Although the Reynolds,
Deborah and capillary numbers describing this flow are all moderate by
laboratory standards, it should be kept in mind that the physical velocities
are relatively large, meters to hundreds of meters per second (depending on
one's interpretation of the Lennard-Jones monomers), and not at all small
for the microscopically thin bodies of liquid considered here.

\section{Constant Velocity Stretching}

We first consider filament stretching at a relatively low constant velocity,
displacing the two ends at $\pm 0.05\sigma/\tau$.  Successive 
snapshots of the filament appear in Fig.~1, in a low-resolution side view
intended to indicate the overall shape.  At early times there is a 
mostly-uniform deformation, aside from some necking near the two ends which
results from the tethering method used here -- the monomers there are 
essentially pinned to their initial relative positions and when these are
moved apart the melt attempts to fill in.  A ``dynamical'' necking down 
appears near the center of the filament around $1500\tau$, followed by 
relatively uniform stretching of the central neck 
region through $3100\tau$, while the two end segments of the filament 
translate with the tethers with only minor changes of shape.  
Eventually the filament breaks when the neck region
is so stretched as to run out of molecules.  As seen in the close-up views 
in Fig.~2, the neck eventually thins down to a few molecular chains lined up
along the stretch axis, and the breakage occurs when these chains are pulled
in opposite direction and slide past each other.  Note that the molecular
model used here does not incorporate chain breaking, so the molecular 
``depletion'' seen here is the only available rupture mechanism.  

Calculations and laboratory experiments \cite{eggers,nagel,basaran} on 
the thinning of a continuum Newtonian liquid cylinder typically find a 
power-law variation of
minimum radius with time. Here, however, while we do observe a systematic 
decrease as seen in Fig.~3, the detailed time variation is not well fit by a 
power of time.  Aside from the fact that the later stages of rupture are 
associated with a neck made of just a small number of highly
extended molecules, it is not unreasonable to see quantitatively different 
behavior.  Furthermore, at this stage the liquid is presumably highly
viscoelastic so that Newtonian fluid predictions may not be applicable. 
Similar behavior is observed in all of the other cases presented below:
although there is a systematic decrease in the filaments' minimum 
radius with time, it does not fit a power-law variation.
Figures 8 and 9 show a sequence of snapshots at higher (fixed) tether speeds, 
$\pm 0.15$ and $\pm 0.5\sigma/\tau$ respectively.  In these various 
simulations, the
initial liquid cylinder is always the same, and within each sequence  
the same length scale is used.  First note that the rupture occurs
at different points along the filament, indicating that the process is 
random and an indirect function of the particular sequence of fluctuating 
positions and velocities in each case, and furthermore that there is no
special defect associated with the ``sample preparation''.  At these higher
speeds, the stagnant end regions at the ends do not develop, and more
molecules are available to the central portion of the filament, allowing
it to attain longer strains before rupture.  Here the ends are steadily
drawn out, leaving a central region which does expand but at a distinctly 
slower speed, and begins to evolve into isolated drops.
In the central region, relatively rapid thinning occurs at several 
intermediate points, and the configuration prior to rupture resembles the 
results of the quasi-static Rayleigh instability \cite{old_rupt}, where 
a cylinder of 
liquid without imposed flow breaks up into spherical drops.  The final stages
of rupture, again, amount to a very narrow neck only a few molecules in
width attenuating down to nothing as its elongated chains are pulled past
each other.
The shape evolution is in a sense opposite to that of the low-velocity case 
in Fig.~1, where the end regions began to translate passively before 
becoming highly stretched.  Finally, as indicated in Table~I, note that 
the maximum length attained
before breakup increases with pulling speed while the time to rupture
decreases, presumably because the higher stress allows more molecules to 
be stretched.

\section{Constant Strain Rate Stretching}

The more common experimental protocol in liquid bridge systems is stretching
at constant strain rate, $\dot{\epsilon}_0 = \dot{L}(t)/L(t)$, leading to an
exponential increase in length and velocity.  Ostensibly the motivation is
to provide a constant ``strain environment'' for the liquid, to facilitate
study of its extensional rheology, although in practice \cite{ms,bkb} the 
strain can often be very unevenly distributed.  In Fig.~10 we show some of 
the time-sequence of shapes observed at a strain rate $\dot{\epsilon}_0 = 
0.001\tau^{-1}$.  The evolution combines features of the various 
constant-velocity
deformations presented above:  the filament narrows in the center and
eventually ruptures there due to attenuation of molecules, but at the same
time the end regions are not translating stagnantly but rather stretch at a
slower rate than the bulk, and at the time of rupture are themselves in 
the process of thinning and presumably would eventually break up due to the 
Rayleigh instability. 

The distinction between constant-velocity and constant-strain deformations
is that in the former case the strain rate (and stress) decrease with time,
so that after an initial stage where the system is pulled out of its
equilibrium configuration, the molecules can relax back;  see Fig.~4.
In the present case, the stress is ostensibly constant, and the molecules are 
increasingly elongated and oriented with no configurational relaxation 
until nearly the time of rupture, as shown in Fig.~11.  The behavior of the
velocity profiles is also consistent with these remarks;  as seen in Fig.~12,   
the velocity is closer to a linear variation along the axis for a
considerably longer time interval. 

A somewhat related MD study \cite{kroger} of the elongation and relaxation 
of a model polymer melt considered chains of 30 monomers using the same 
interactions as in this paper (but with a short $r_c=2^{1/6}\sigma$ LJ 
cutoff).  Instead of simulating the explicit stretching of filaments, a 
nonequilibrium MD method was used to apply a constant-strain uniaxial 
elongation to a homogeneous system in 
a box.  Among other things, the behavior of the chain size (radius of
gyration) and orientation was measured, and the qualitative behavior is 
similar to that seen here.  The elongation and degree of orientation
increased monotonically during the flow, at a rate which increased
monotonically with the strain rate, and then relaxed monotonically when
the flow was turned off.  

We have considered the question of unrecoverable strain \cite{mp} in this 
simulation, by releasing the tethered monomers and allowing the liquid to take
up a new equilibrium shape without external stress.  If the filament is 
released well
before rupture, it seems to withdraw into a sphere, whereas if released 
just before rupture the thinning neck continues to thin, rupture occurs 
anyway, and afterwards the liquid appears to be withdrawing into two
spheres.  Here the phrase ``appears to be'' means that as a function of time
the liquid shape is tending towards one or two spheres at the original 
liquid density, but the process is so slow that we have not followed it 
all the way to 
equilibrium.  The asymptotic strain does, however, go well below 1/2,
and we do not see the values of unrecoverable strain suggested in \cite{mp}  
Similar behavior is found in the other cases as well.  Given that the
initial simulated filament was prepared by an equilbration procedure
involving relaxation of a dilute high-temperature melt, it is difficult to
see how the final state here would differ in any statistically significant 
way, other than perhaps the time needed to reach it.

At higher strain rates a different behavior appears, a cohesive failure of
the melt in the region where the molecules are tethered to pulling sites.
In Fig.~13, at strain rate $\dot{\epsilon}_0 = 0.004$, there is a rapid
thinning at the ends of the filament leaving a central stagnant region 
showing the onset of a Rayleigh instability, while the thin necks attaching
this region to the ends attenuate until breakup.  At the same time, some of 
the tethered molecules are pulled out of the filament, and the fluid bodies
at the ends have not had time to relax to an equilibrium shape.
At the still higher strain rate $\dot{\epsilon}_0 = 0.01$ shown in Fig.~14, 
the tethered molecules are simply pulled free of rest of the the filament,
and rapid rupture occurs.  The rupture times and lengths for the
various cases are recorded in Table~I.

\section{Force Measurements}

As an alternative to the stress fields, we have studied the forces from a 
more direct viewpoint on the monomer scale. The quantity of perhaps
most direct experimental relevance is the pulling force exerted on the 
tethers at the ends.  While this quantity is easy to evaluate in the
simulations, it is unfortunately quite noisy.
Aside from the problem noted above of large atomic-level force fluctuations
in general, here the procedure used to generate the motion exacerbates the
problem.  The atoms at the ends of the filament are attached to moving tether 
sites by stiff springs, whose force is very sensitive to small atomic
displacements.  

In the constant-velocity simulations the result is 
simple -- the pulling force is roughly constant in time, fluctuating about 
a value which is approximately linearly proportional to
the pulling velocity.   At the three velocities studied, 0.05, 0.15 and 
0.5$\sigma/\tau$, the force per pulled monomer is measured to be 2.72$\pm$0.16, 
7.31$\pm$0.24 and 24.9$\pm$0.25$\epsilon/\sigma$, respectively.  (Typical
values of the instantaneous force on an monomer at this density and temperature
are O(1-10) in this unit, while its time averaged values over a 100$\tau$
interval are an order of magnitude smaller.)  One can rationalize the
time-constancy of the force by arguing that in this case the molecules are 
moving
past each other with a roughly fixed value of relative velocity, and since
the material is a viscous liquid in a low Reynolds number flow, there is a
constant drag force on each molecule.   

In contrast, in the constant strain rate cases there is a non-trivial time
variation.  For example, as shown in Fig.~15 for the lowest rate 
$\dot{\epsilon}_0=0.001\tau^{-1}$, the tether force jumps to a finite value
once the force is applied, then increases with time until about 1750$\tau$.
After a sharp drop, the force again increases until rupture.
Referring to Fig.~10, after this transition time the right and left sides of 
the filament are approximately translating without a significant shape change,
while the central region thins down to nothing.  The early increase can then
be associated with strain hardening \cite{ms,bkb} and the shape deformation 
of the entire 
liquid filament; the sharp drop with relaxation of stress after some internal
readjustment and formation of the central neck; the second increase 
with the strain hardening of the central thinner region; and the final
drop-off with rupture.  At the next highest strain value, 0.004$\tau^{-1}$,
the numerical values of the tether force are comparable, and the tether
force shows a similar variation, with a corresponding drop at about
600$\tau$, the time at which (see Fig.~13) the central region of the
filament halts and ceases to change its shape while the regions to either 
side continue to elongate.  At still highest strain rates there is no obvious
pattern to the tether force evolution;  the fragmentation of the regions
near the tethers makes it difficult to identify simple mechanisms.

We can also compute the forces in the interior of the filament, and in 
Fig.~16 we show an example of the force in the axial direction in the 
$\epsilon_0=0.001$ constant strain rate simulation at time 2150$\tau$, 
averaged over a
100$\tau$ time interval and a 6$\sigma$ spatial interval along the axis.  
The force on the end regions containing the tethered monomer shows
large spatial and temporal fluctuations for the reasons given at the
beginning of this Section, and these two regions have been suppressed in 
this plot. Aside from the fluctuations, the trend is a roughly-linear ramp 
with positive (negative) values corresponding to the halves of the filament 
moving to the right (left).   As a function of time, the fluctuations shift
around apparently randomly, and the height of the ramp is roughly constant in 
time.  At higher strain rates, the analogous plot is again a fluctuating
quasi-linear ramp in its central region, but shows rapid spatial fluctuations 
near the ends, corresponding to molecules being pulled out there.  In 
contrast, in the constant-velocity simulations there is always a linear ramp
as a function of $z$, and furthermore, at different pulling speeds the ramp 
height is roughly linear in velocity.  

\section{Conclusions}

We have discussed the results of molecular simulations of a liquid filament
of model polymer melt stretched to the point of rupture.  These model systems
are composed of non-breakable, freely-jointed, entangled chains, and the 
rupture process
typically involves ductile failure of some sort.  Here we observed these 
filaments stretch into a very non-uniform shape, with one or more thin necked 
regions, and then fail when one of the necks simply attenuates down to
nothing.  More precisely, during the stretching process the thinning neck 
regions are populated with well-extended chain molecules lying roughly 
parallel to the extension axis, which
are drawn past each other to one side or the other by their respective 
interacting neighbors.  Eventually the filament is held together by only a 
few such adjacent elongated chains, and splits once these slide past each 
other.  In continuum terms, the neck thins under a combination of mass 
transfer due to outflow and (Rayleigh) capillary instability.
In analogous earlier simulations in systems of melts or solutions based on
shorter length molecules (up to 20 monomers), presumably unentangled, a more 
familiar form of ductile fracture was observed in which a neck several 
molecules in thickness separated into two pieces \cite{bkb}.  
In that case, a model solution was able to attain the longest lengths prior 
to breakup, and there one could see that the fully-extended longer 
molecules acted as a backbone which stiffened the filament, and breakage
occurred when the filament was extended to the point that these backbone
molecules were no longer in contact with each other.  

In term of the classification of failure zones by Malkus and Petrie 
\cite{mp}, the constant velocity and lowest-rate constant strain runs are in
the ``flow zone'', where viscous deformation can procede indefinitely 
provided there is enough liquid.  The two runs at higher constant strain
appear to have ``glassy states'', at least near the tethers where rupture
occurs.

The filament shapes observed here seem unusually irregular compared to those 
studied elsewhere, and in no case do we observe a long neck of nearly constant 
diameter.  Note however, that these simulations begin with a liquid sample 
whose diameter is already in the nanometer range, and so these results should 
at best be compared to the last, usually poorly-quantified stages of typical 
laboratory studies.  The
surface irregularities and fluctuations are, however, quite characteristic 
of molecular simulations of moving microscopic liquid bodies with free 
surfaces \cite{old_rupt,uzi}.  In consequence of these fluctuations, the 
measured continuum fields show sufficiently large statistical fluctuations 
as to be non-trivial to understand even qualitatively.  Even microscopic 
measurements which do yield a smooth result, such as the variation of minimum
filament radius with time, are not in agreement with continuum calculations.  
The molecular level information is likewise fluctuating, but does seems to 
be readily interpretable at this scale. 

A further issue in comparing these results to laboratory measurements 
is the pulling speed involved.  Although the standard dimensionless  
groups characterizing these flows, the Reynolds, Deborah and capillary 
numbers, are similar to those considered elsewhere, there is quite a 
difference in velocities. In typical liquid flows the drift or peculiar 
velocity is a small fraction of the thermal velocity of the monomers, but
here, particularly in the late stages of the constant strain simulations,
the tether velocity approaches the thermal velocity, O(1$\sigma/\tau$).
The severe fragmentation of the end regions of the filament is presumably 
one aspect of this high speed, but a second is that surface tension effects
which would ordinarily smooth a free surface may not have time to act.

\section*{Acknowledgments}  We thank 
the NASA Office of Physical and Biological Sciences for financial support, 
and the NASA Advanced Supercomputing Division at the Ames Research Center, 
and NPACI at the San Diego Supercomputer Center for providing computational 
resources.

\vspace*{1.0in}

\begin{center}

\begin{tabular}{|c||c|c|c||c|c|c|}\hline
        & \multicolumn{3}{c||}{\, Constant velocity\,}
        & \multicolumn{3}{c|}{\, Constant strain rate\,} \\ \hline\hline
$v_0$ or $\dot{\epsilon}_0$ & 0.05 & 0.15 & 0.5 & 0.001 & 0.004 & 0.01\\ \hline
length & 432 & 694 & 1080 & 701 & 1594 & 959 \\ \hline
time & 3520 & 2045 & 1000 & 2250 & 763 & 263 \\ \hline
\end{tabular}

\vspace{0.5in}
{\bf Table I}. Final length and time to rupture (MD units). 

\end{center}

\newpage

\begin{figure}
\epsfxsize=5.0in\epsfysize=5.0in
\epsfbox{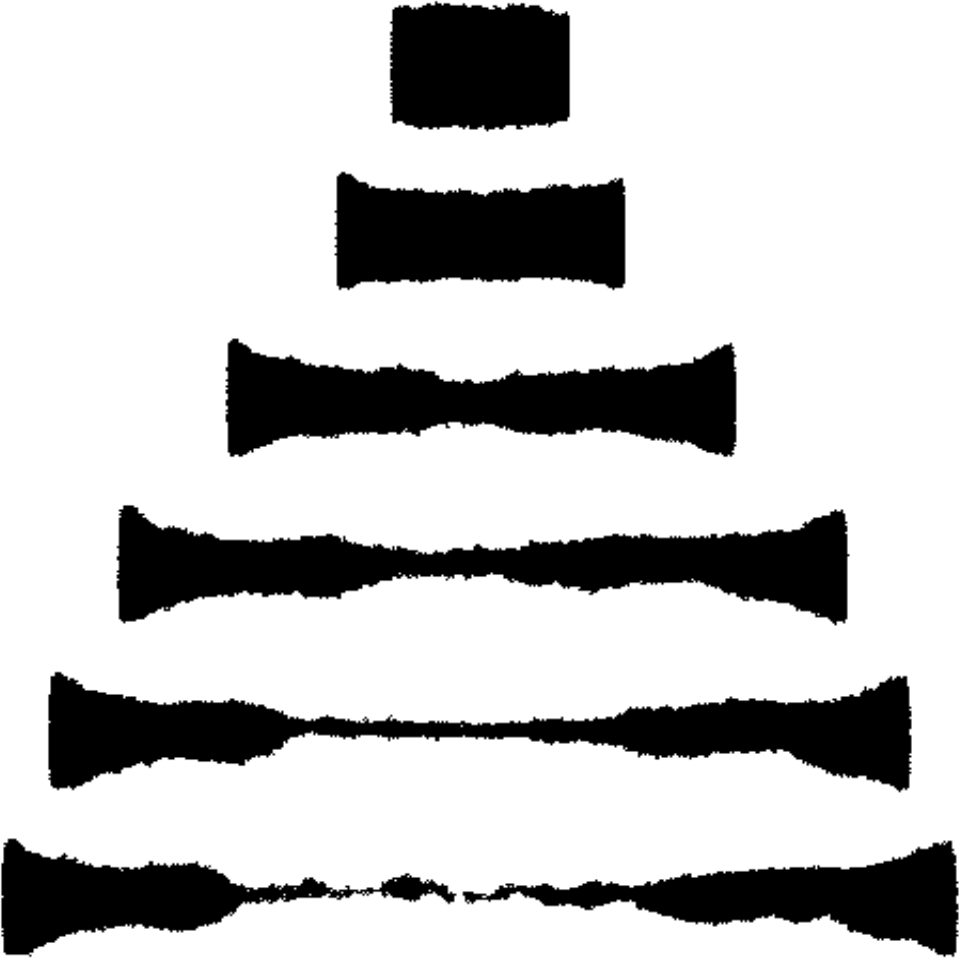}
\vspace{1cm}
\caption{Sequence of filament shapes in a constant velocity deformation at
pulling speed $v_0=0.05\sigma/\tau$, at times 0, 500, 1500, 2500, 3100 and
3520$\tau$ (top to bottom).}
\end{figure}

\newpage

\begin{figure}
\epsfxsize=5.0in\epsfysize=3.5in
\epsfbox{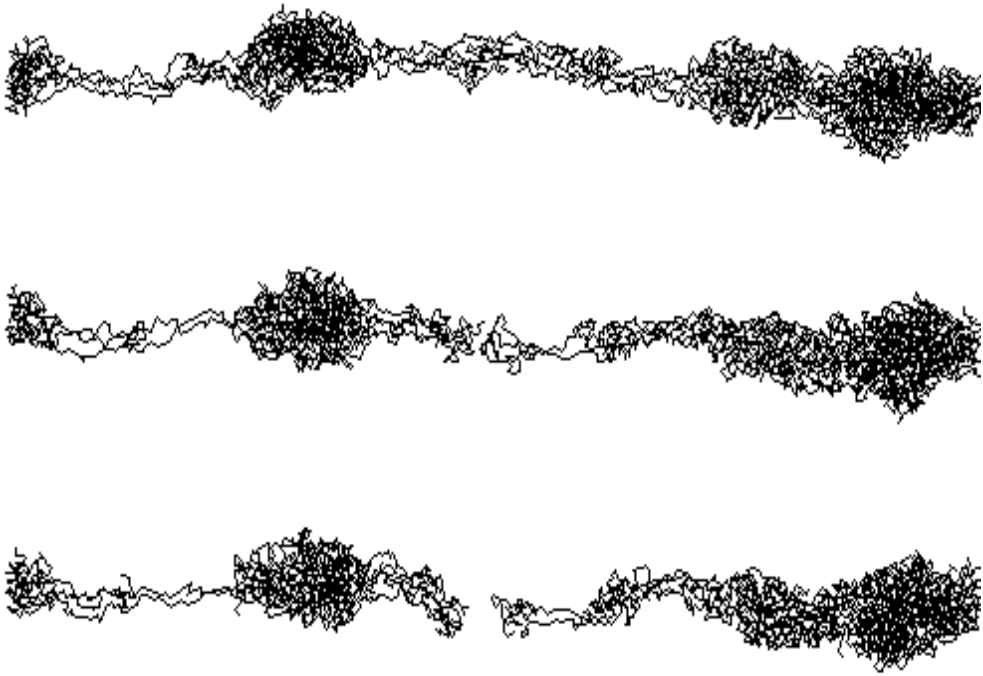}
\vspace{1cm}
\caption{Rupture region in the constant velocity deformation of Fig.~1, 
at times 3475, 3515 and 3520$\tau$ (top to bottom).}
\end{figure}

\vspace{0.5in}

\begin{figure}
\epsfxsize=3.0in\epsfysize=3.0in
\epsfbox{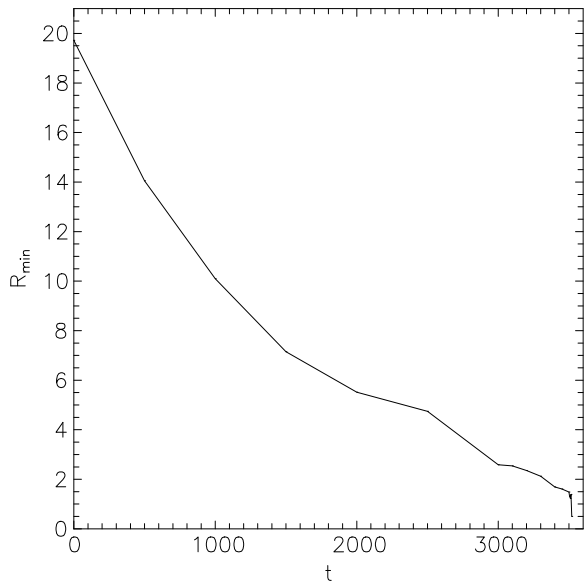}
\vspace{1cm}
\caption{Minimum radius {\em vs}. time in the constant velocity 
deformation of Fig.~1.}
\end{figure}

\newpage

\begin{figure}
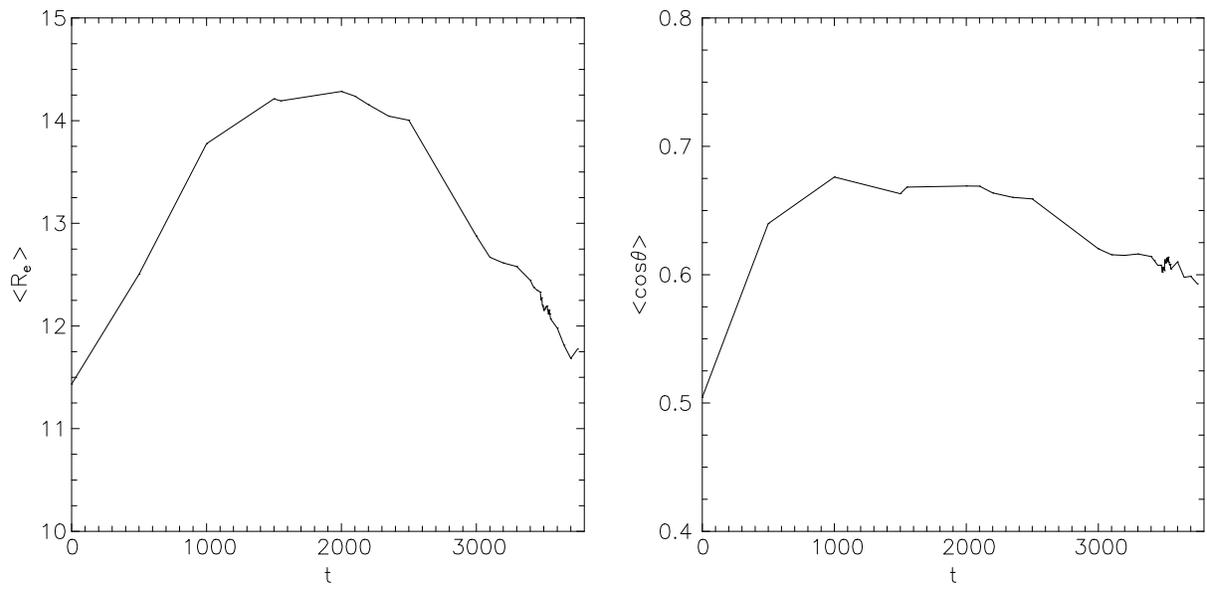

\centerline{
\hbox{
\epsfxsize=3.0in\epsfysize=3.0in
\epsfbox{fig4left.epsi}
\hspace{.5cm}
\epsfxsize=3.0in\epsfysize=3.0in
\epsfbox{fig4right.epsi} } }
\vspace{1cm}
\caption{Mean molecular end-to-end length and orientation
angle {\em vs}. time in the constant velocity deformation of Fig.~1.} 
\end{figure}

\newpage

\begin{figure}
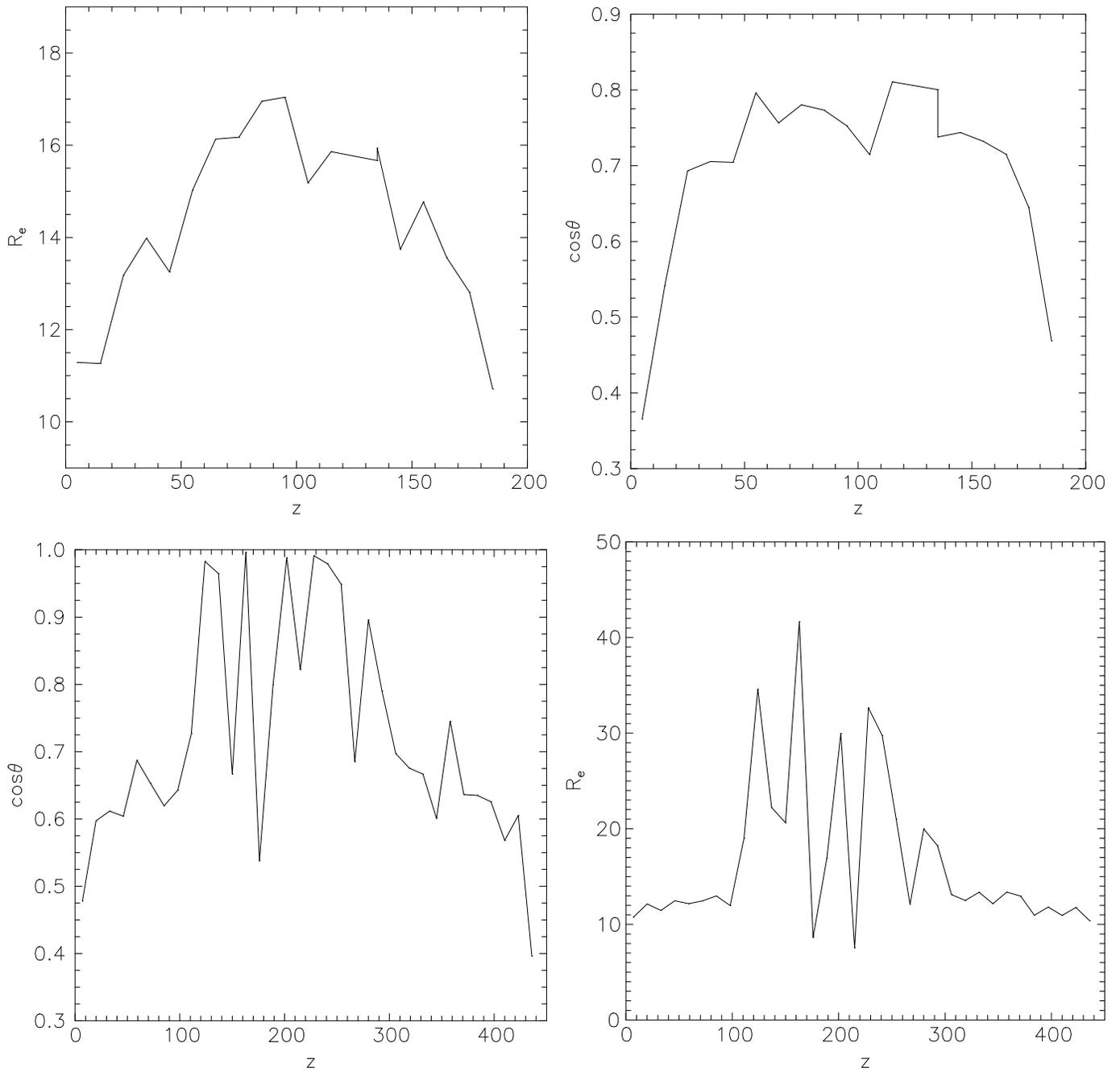

\centerline{
\hbox{
\epsfxsize=3.5in \epsfbox{fig5topleft.epsi}
\hspace{.2cm}
\epsfxsize=3.5in \epsfbox{fig5topright.epsi} } }
\vspace{.3cm}
\centerline{
\epsfxsize=3.5in \epsfbox{fig5botleft.epsi}
\hspace{.2cm}
\epsfxsize=3.5in \epsfbox{fig5botright.epsi} }
\vspace{.1cm}
\caption{Molecular length and molecular orientation {\em vs}. axial
position, in the constant velocity deformation of Fig.~1.
(top) at time 1000$\tau$, (bottom) at time 3500$\tau$.} 
\end{figure}

\newpage

\begin{figure}
\centerline{
\hbox{
\epsfxsize=3.5in \epsfbox{fig6topleft.epsi}
\hspace{.2cm}
\epsfxsize=3.5in \epsfbox{fig6topright.epsi} } }
\vspace{.3cm}
\centerline{
\epsfxsize=3.5in \epsfbox{fig6botleft.epsi}
\hspace{.2cm}
\epsfxsize=3.5in \epsfbox{fig6botright.epsi} }
\vspace{.1cm}
\caption{Velocity {\em vs}. axial position in the constant 
velocity deformation of Fig.~1 at the times indicated.
Each curve represents a 50$\tau$ time average.}
\end{figure}

\newpage

\begin{figure}
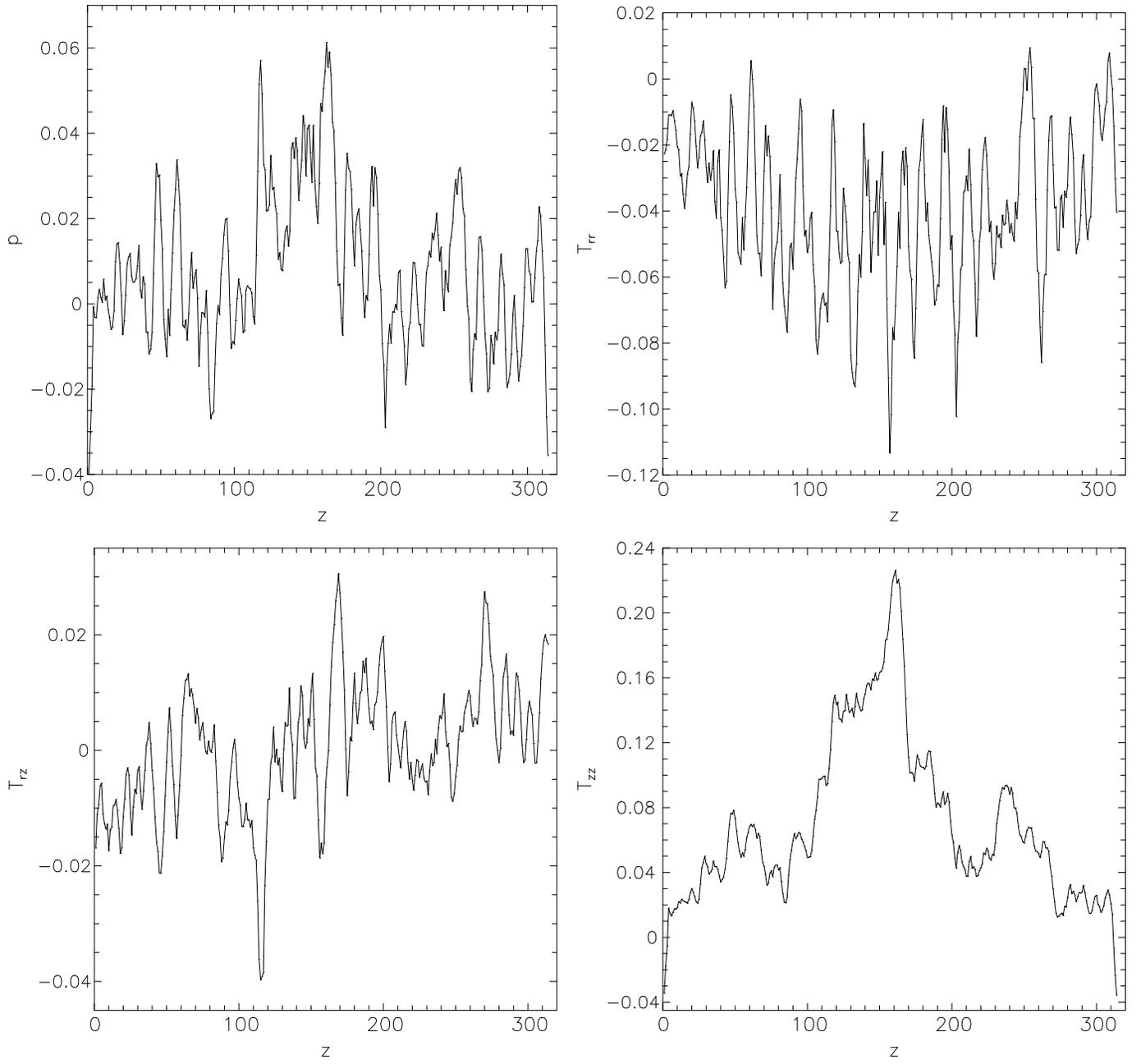

\centerline{
\hbox{
\epsfxsize=3.5in \epsfbox{fig7topleft.epsi}
\hspace{.2cm}
\epsfxsize=3.5in \epsfbox{fig7topright.epsi} } }
\vspace{.3cm}
\centerline{
\epsfxsize=3.5in \epsfbox{fig7botleft.epsi}
\hspace{.2cm}
\epsfxsize=3.5in \epsfbox{fig7botright.epsi} }
\vspace{.1cm}
\caption{Stress tensor components {\em vs}. axial position in the constant
velocity deformation of Fig.~1 at time 2500$\tau$. Clockwise from upper
left: pressure ($p$), radial ($rr$), axial ($zz$), and shear ($rz$) stress.}
\end{figure}

\newpage

\vspace*{-1.0in}

\begin{figure}
\epsfxsize=5.0in\epsfysize=5.0in
\epsfbox{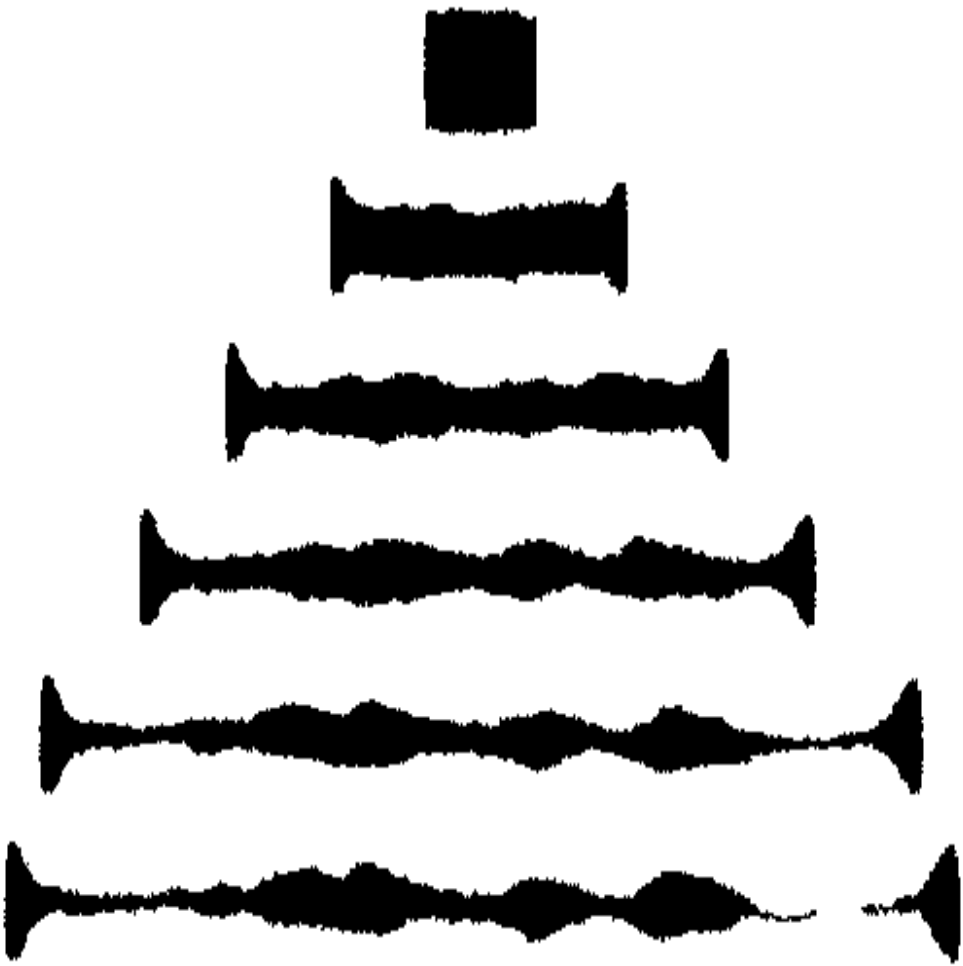}
\vspace{1cm}
\caption{Sequence of filament shapes in a constant velocity deformation at 
pulling speed $v_0=0.15\sigma/\tau$, at times 0, 450, 950, 1770, 1870 and
2045$\tau$.}
\end{figure}

\vspace*{-1.0in}

\begin{figure}
\epsfxsize=5.0in\epsfysize=5.0in
\epsfbox{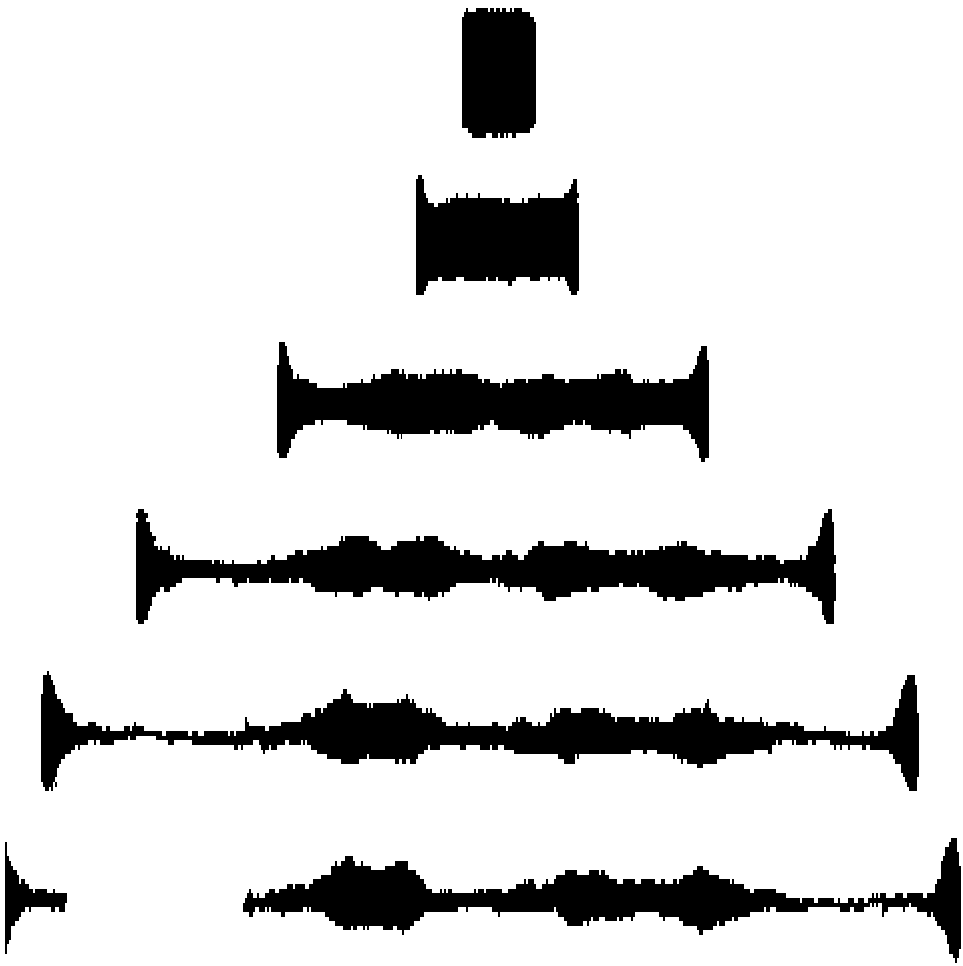}
\vspace{1cm}
\caption{Sequence of filament shapes in a constant velocity deformation at
pulling speed $v_0=0.5\sigma/\tau$, at times 0, 100, 400, 700, 900 and
1000$\tau$.}
\end{figure}

\newpage

\begin{figure}
\epsfxsize=5.0in\epsfysize=5.0in
\epsfbox{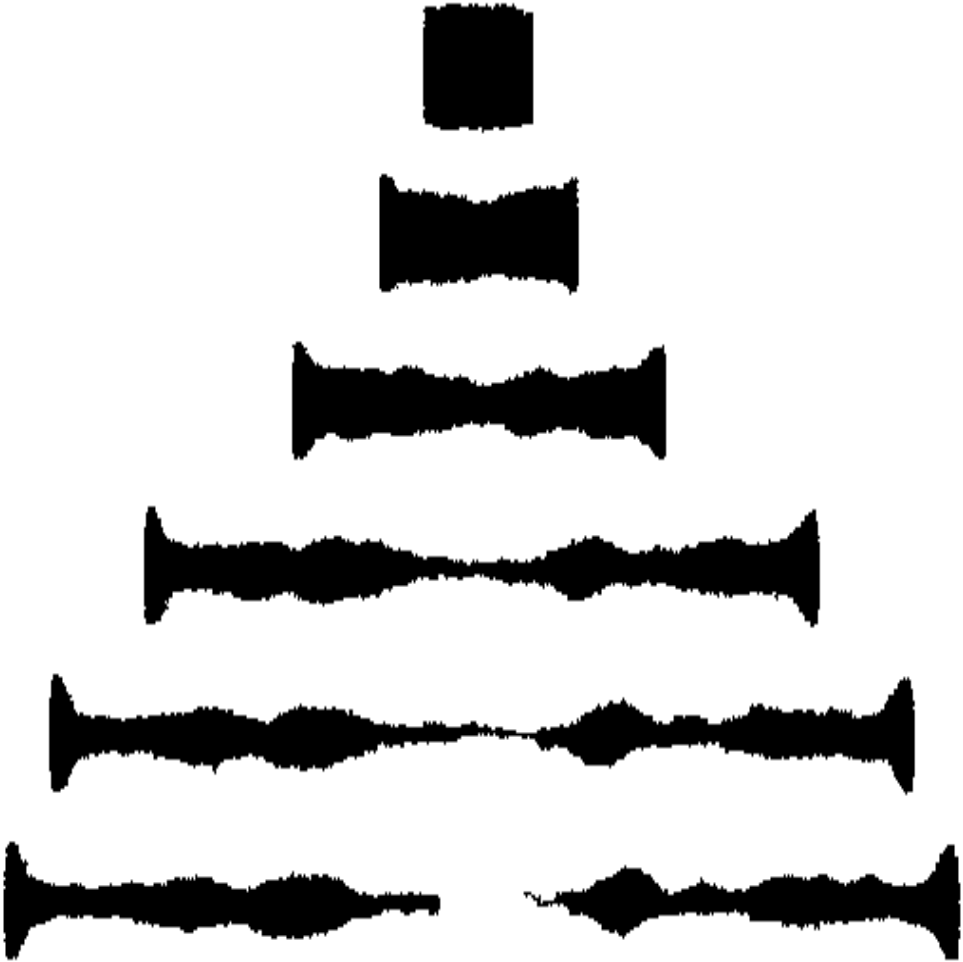}
\vspace{1cm}
\caption{ Sequence of filament shapes in a constant strain rate deformation at 
$\dot{\epsilon}_0=0.001\tau^{-1}$, at times 0, 650, 1300, 1950, 2150 and 
2250$\tau$.}
\end{figure}

\newpage

\begin{figure}
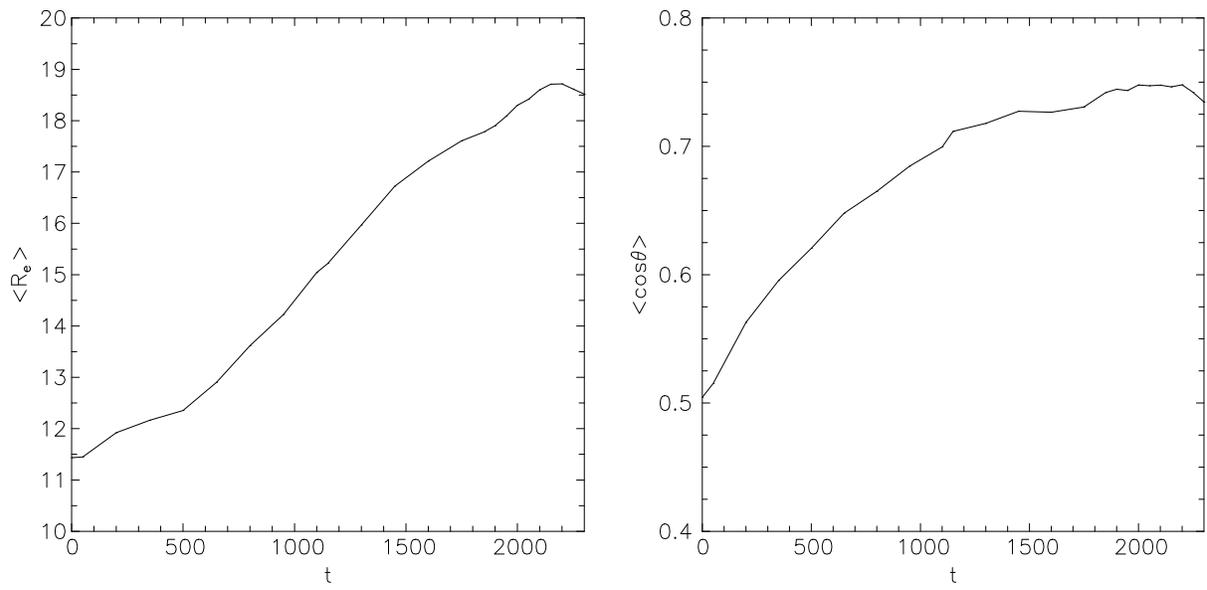

\centerline{
\hbox{
\epsfxsize=3.0in\epsfysize=3.0in
\epsfbox{fig11left.epsi}
\hspace{.5cm}
\epsfxsize=3.0in\epsfysize=3.0in
\epsfbox{fig11right.epsi} } }
\vspace{1cm}
\caption{Mean molecular end-to-end length and orientation
angle {\em vs}. time in the constant strain deformation of Fig.~10.}
\end{figure}

\newpage

\begin{figure}
\centerline{
\hbox{
\epsfxsize=3.5in \epsfbox{fig12topleft.epsi}
\hspace{.2cm}
\epsfxsize=3.5in \epsfbox{fig12topright.epsi} } }
\vspace{.3cm}
\centerline{
\epsfxsize=3.5in \epsfbox{fig12botleft.epsi}
\hspace{.2cm}
\epsfxsize=3.5in \epsfbox{fig12botright.epsi} }
\vspace{.1cm}
\caption{Velocity {\em vs}. axial position in the constant 
velocity deformation of Fig.~10 at the times indicated.
Each curve represents a 50$\tau$ time average.}
\end{figure}

\newpage

\vspace*{-1.0in}

\begin{figure}
\epsfxsize=5.0in\epsfysize=5.0in
\epsfbox{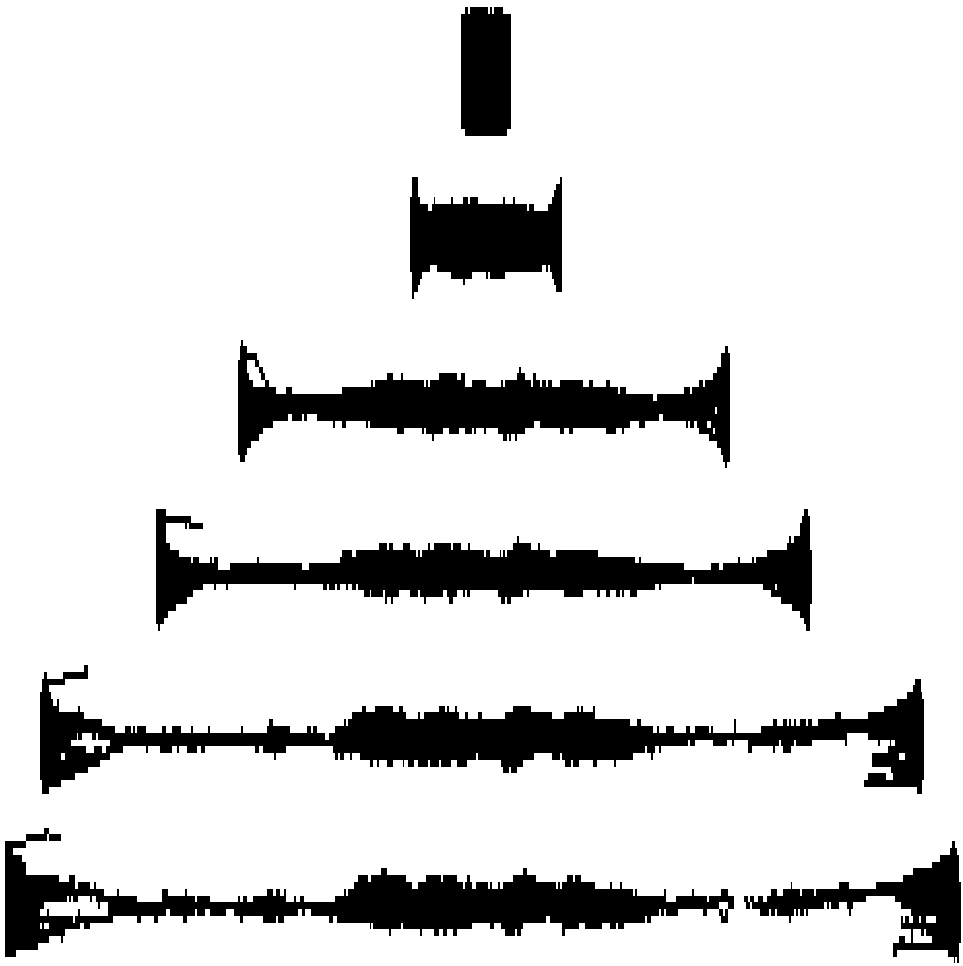}
\vspace{1cm}
\caption{ Sequence of filament shapes in a constant strain rate deformation at
$\dot{\epsilon}_0=0.004\tau^{-1}$, at times 0, 295, 595, 668, 743 and
763$\tau$.}
\end{figure}

\vspace*{-1.0in}

\begin{figure}
\epsfxsize=5.0in\epsfysize=5.0in
\epsfbox{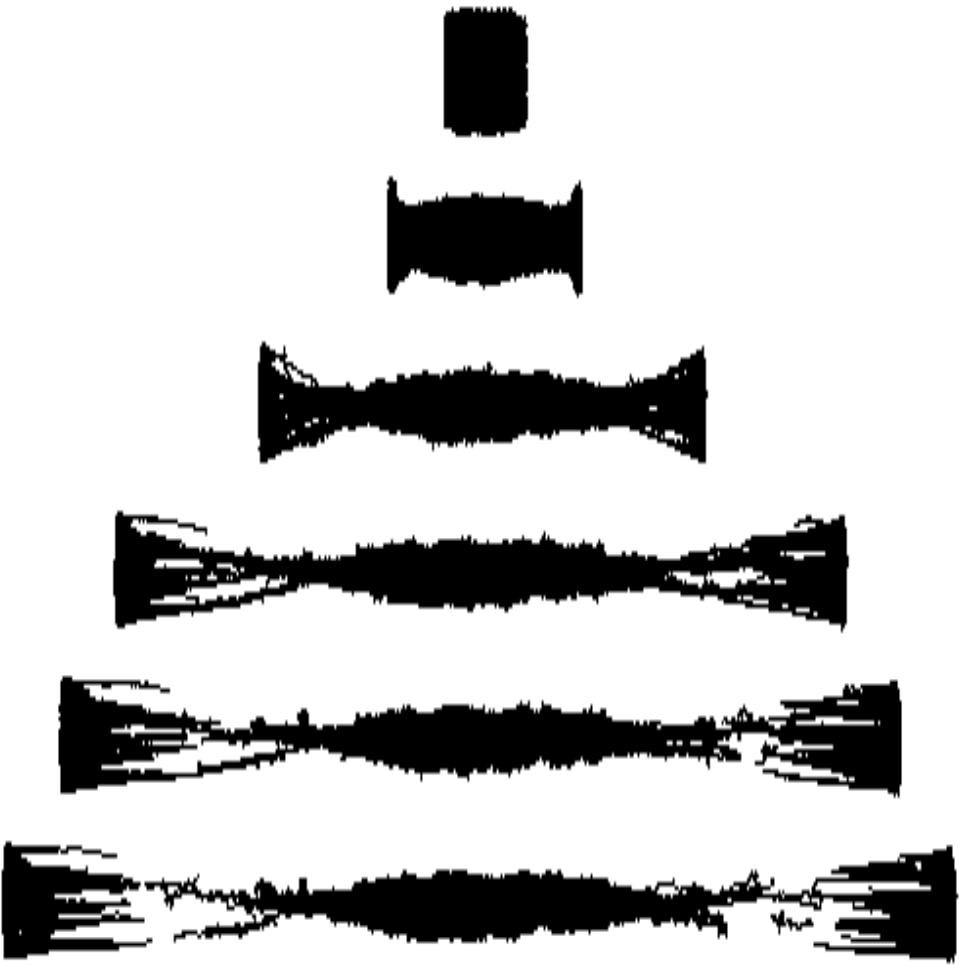}
\vspace{1cm}
\caption{Sequence of filament shapes in a constant strain rate deformation at
$\dot{\epsilon}_0=0.01\tau^{-1}$, at times 0, 90, 175, 225, 240 and
253$\tau$.}
\end{figure}

\newpage

\begin{figure}
\epsfxsize=3.0in\epsfysize=3.0in
\epsfbox{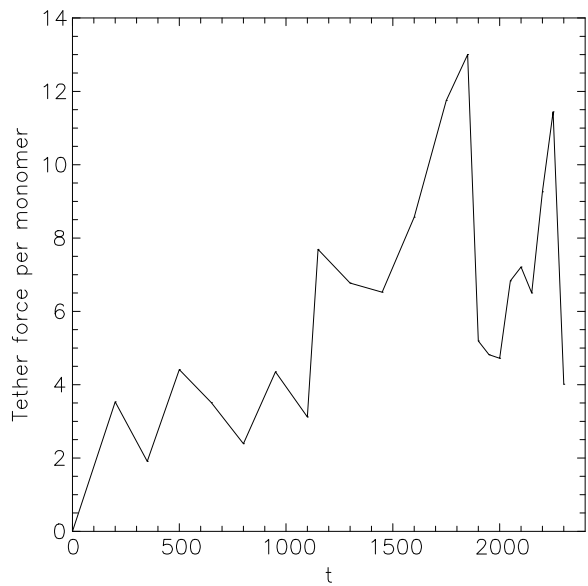}
\vspace{1cm}
\caption{ Variation of the tether force with time, in the simulation sshown in
Fig.~10.}
\end{figure}

\begin{figure}
\epsfxsize=3.0in\epsfysize=3.0in
\epsfbox{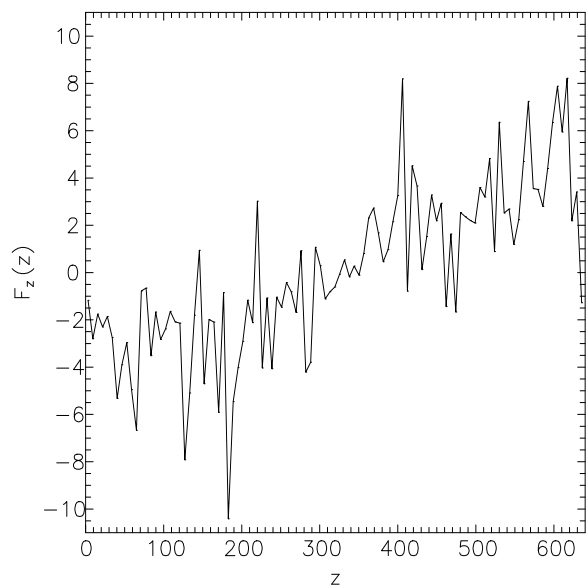}
\vspace{1cm}
\caption{ Axial force as a function of axial position in the filament shown
in Fig.~10, at time 2150$\tau$.}
\end{figure}

\end{document}